\documentclass[aps,pre,amsmath,amssymb,twocolumn]{revtex4}
\usepackage{amsmath,amsthm,amssymb}
\allowdisplaybreaks
\usepackage{graphicx}
\usepackage{epstopdf}
\usepackage{pstricks}
\usepackage{epsf}

\theoremstyle{plain}

\usepackage{graphicx}
\newif \ifLastSection \LastSectionfalse

\numberwithin{equation}{section}

\newcommand{\be}{\begin{equation}}
\newcommand{\ee}{\end{equation}}
\newcommand{\ba}{\begin{eqnarray}}
\newcommand{\ea}{\end{eqnarray}}
\newcommand{\baa}{\begin{eqnarray}}
\newcommand{\eaa}{\end{eqnarray}}
\newcommand{\ed}{\end{document}}

\newcommand{\re}[1]{(\ref{#1})}
\newcommand{\ci}[1]{\cite{#1}}

\begin{document}

\title {Thermal diffusion in branched structures: Metric graph based approach}
\author{K. Sabirov$^{a,b}$, Zh. Zhunussova$^c$, D. Babajanov$^a$ and D. Matrasulov$^a$}
\affiliation{ $^a$ Turin Polytechnic University in Tashkent, 17
Niyazov Str.,
100095,  Tashkent, Uzbekistan\\
$^b$ Tashkent University of Information Technologies, 108 Amir Temur Str., 100200, Tashkent Uzbekistan\\
$^c$ Al-Farabi Kazakh National University, 71 Al-Farabi avenue,
400050, Almaty, Kazakhstan}

\begin{abstract}
We consider the problem of heat diffusion in branched systems and
networks on the basis of a model described in terms of the heat
equation on metric graphs. Using the explicit analytical solutions
of the latter, evolution of the temperature profile and heat flow
on each branch are computed. Extension of the study for nonlinear
regime is considered using a nonlinear heat equation on metric
graphs. It is found that in nonlinear regime is more intensive
than that in linear case.
\end{abstract}
\maketitle

\section{Introduction}
Particle and wave transport in low dimensional branched structures
is of importance for different technological applications.
Discrete structures, networks and other types of branched systems
appear as part of many materials and devices at macro-, micro- and
even nanoscale.  Tuning, optimization and controlling
functionality of such devices require deep understanding of wave
phenomena in such branched structures and effective, maximally
realistic modeling of these phenomena. The latter requires using
simple and powerful approaches for such modeling. One of such
approaches is based on the use of different wave (partial
differential) equation on so called metric graphs. The  graph
itself is determined as a set of bonds connected to each other at
the vertices (branching points) according to some rule which is
called the topology of a graph. When bonds of a graph are assigned
length it is called metric graph. Topology of a graph is given in
terms of the adjacency matrix which can be written as
\cite{Uzy1,Uzy2}:
\begin{eqnarray}
C_{ij}=C_{ji}=\left\{\begin{array}{ccl}&1 & \mbox{ if }\; i\;
\mbox{ and }\; j\; \mbox{ are connected, }\\
& 0 & \mbox{ otherwise, }\end{array}\right. \\
i,j=1,2,...,V.\nonumber
\end{eqnarray}

Study of wave dynamics in branched systems using metric graphs can
be divided into two cases:  transport of the waves in linear and
nonlinear regimes. Investigation of the linear waves in branched
systems are focused mainly on so-called "quantum graphs" which are
described in terms of linear Schrodinger equation on metric
graphs. Pioneering treatments of quantum mechanical motion in
branched systems has been considered few decades ago in the Refs.
\ci{Pauling}-\ci{Alex}. However, strict treatment of quantum
graphs was first presented by Exner, Seba and Stovicek to describe
free quantum motion on branched wires \ci{Exner1}. Later Kostrykin
and Schrader derived the general boundary conditions providing
self-adjointness of the Schr\"odinger operator on graphs
\ci{Kost}. Bolte and Harrison extended such boundary conditions
for the Dirac operator on metric graphs \ci{Bolte}. Hul \emph{et
al} considered experimental realization of quantum graphs in
optical microwave networks \ci{Hul}. An important topic related to
quantum graphs was studied in the context of quantum chaos theory
and spectral statistics \ci{Uzy1,Bolte,Uzy2,Uzy3,Harrison}.
Spectral properties and band structure of periodic quantum graphs
also attracted much interest \ci{Grisha1,Kotlyarov}. Different
aspects of the Schr\"odinger operators on graphs have been studied
in the Refs.\ci{Exner15,Grisha,Mugnolo,Rabinovich,Bolte1}. Despite
great progress made on the study of quantum graphs, other wave
equations on metric graphs are still remaining out of the focus,
although some  mathematical aspects of partial differential
equations on graphs have been considered in the literature (see,
e.g., \ci{PDEr1}- \ci{HEBook5}). Nonlinear wave dynamics in
branched systems which are described by nonlinear evolution
equations on metric graphs has attracted much attention during
past decade \ci{Zarif} -\ci{Karim2018}.

In this paper we consider dynamics of thermal waves in branched
systems which are modeled in terms of the linear and nonlinear
heat equations on metric graphs. Such a model can be used for
describing the classical (non-quantum) heat diffusion in branched
polymers, carbon nanotube networks and any low-dimensional
branched structures appearing in condensed matter physics.

\begin{figure}[th!]
\includegraphics[width=70mm]{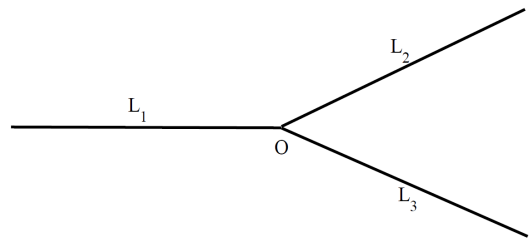}
\vspace{1mm} \caption{Metric star graph. $L_j$ is the length of
the $j$th bond with $j=1,2,3$.} \label{pic1}
\end{figure}

\section{Linear heat diffusion in networks}

Our aim is modeling the heat diffusion in branched structures
using heat equation on metric graphs. Solution of the latter  can
be constructed using that of heat equation on a finite interval.
Therefore before formulating the problem for metric graph, we will
briefly recall solution of the heat equation on a finite interval,
$(0;l)$ which is given by  \ci{HEBook1,HEBook2,HEBook3,HEBook4}

\begin{figure}[th!]
\includegraphics[width=90mm]{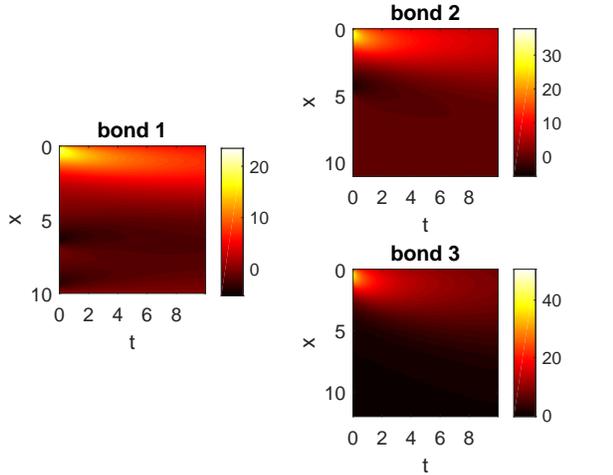}
\vspace{1mm} \caption{(Color online) Evolution of the temperature
 profile on each bond of the
star graph described by Eqs.\re{eq10}-\re{eq14} for the values of
heat conductivity $\kappa_1=0.3,\, \kappa_2=0.5,\, \kappa_3=0.7$
and bond lengths $L_1=10,\, L_2=11,\, L_3=12.$} \label{tprof1}
\end{figure}

\begin{equation} \frac{\partial u}{\partial t}=\kappa^2\frac{\partial^2 u}{\partial x^2},\,x\in (0,l), \label{eq1}
\end{equation}
where $u(x,t)$ is the temperature profile, $\kappa$ is the heat
conductivity. The boundary and initial conditions are imposed as
\begin{equation}
 u(0,t)=u(l,t)=0. \label{eq2} \end{equation}
and
\begin{equation}
 u(x,0)=f(x). \label{eq3}\end{equation}

Exact solutions of the problem \re{eq1}-\re{eq3} can be obtained
by separating variables as
\begin{equation} u(x,t)=X(x)T(t), \label{eq4} \end{equation}
that gives the general solution of Eq.\re{eq1}
\begin{equation}
u(x,t)=\sum_{n=1}^{\infty}C_{n}e^{-\frac{\pi^{2}\kappa^2n^{2}}{l^{2}}t}\sin(\frac{\pi
n}{l}x) \label{eq9}\end{equation}

In the following this solution will be used  to construct the
solution of the heat equation on metric graphs with finite bonds.
We note that mathematical aspects of the heat equation on graphs
have been earlier considered in the Refs.
\ci{HEBook5,HEGr1,HEGr2,HEGr3,HEGr4,HEGr5}. Here we will obtain
explicit solution of heat equation on metric graph and apply it
for modeling heat diffusion in networks.

\begin{figure}[th!]
\includegraphics[width=90mm]{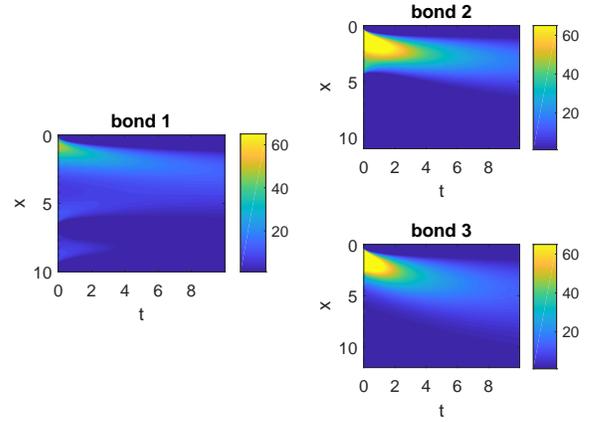}
\vspace{1mm} \caption{(Color online) Heat flux on each bond of the
star graph described by Eqs.\re{eq10}-\re{eq14} for the values of
parameters $\kappa_1=4.5,\, \kappa_2=8,\, \kappa_3=8.4$ and
$L_1=10,\, L_2=11,\, L_3=12$.} \label{hf1}
\end{figure}

Consider the star graph consisting of three bonds $b_{j}
(j=1,2,3)$ with the lengths $L_{j} (j=1,2,3)$ connected at one
vertex (See, Fig.\ref{pic1}). On each bond of this graph we have
heat equation given by
\begin{equation} \frac{\partial u_j}{\partial t}=\kappa_j^2\frac{\partial^2 u_j}{\partial x^2},
\label{eq10}\end{equation}
 with $\kappa_j$ being the heat conductivity of each branch of
 the  graph.
Initial  condition for Eq.\re{eq10} can be imposed as
\begin{equation}
 u_{j}(x,0)=f_{j}(x).  \label{eq11}\end{equation}
To solve Eq.\re{eq10}, one needs to impose also the  boundary
conditions at the branching point (vertex) of graph and at the
bond edges. At the bonds edges we can impose Dirichlet boundary
conditions given by
\begin{equation}
 u_{j}(L_{j})=0,
\label{eq12}\end{equation} while  the vertex boundary conditions
can be imposed as
\begin{equation}
u_{1}(0,t)=u_{2}(0,t)=u_{3}(0,t) \label{eq13}\end{equation}
\begin{equation}
\left.\left(\frac{\partial u_{1}}{\partial x}+\frac{\partial
u_{2}}{\partial x}+ \frac{\partial u_{3}}{\partial
x}\right)\right|_{x=0}=0. \label{eq14}\end{equation}

Eq.\re{eq13} is the continuity of solution at the vertex, while
Eq.\re{eq14} provides Kirchhoff rule for the heat flux. Using the
prescription, proposed in \ci{Kost}, one can easily make sure that
these boundary conditions provide self-adjointness of the heat
equation on a graph.

Variables in Eq.\re{eq10} can be separated by the substitution
\begin{equation}
u_{j}(x,t)=X_{j}(x)T(t), j=1,2,3, \label{eq15}\end{equation} that
gives rise to equations
\begin{equation}
 X_{j}^{\prime\prime}(x)+\frac{\lambda^{2}}{\kappa_j^2}X_{j}(x)=0,  \label{eq16}\end{equation}

\begin{figure}[th!]
\includegraphics[width=70mm]{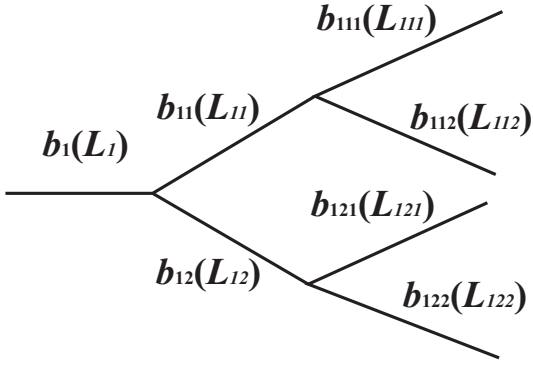}
\vspace{3mm} \caption{Tree graph. $x_1\in [0,L_1], x_{1i} \in
[0,L_{1i}]$, and $x_{1ij} \in [0,L_{1ij}]$ with $i=1,2;j=1,2$.}
\label{tree}
\end{figure}

\begin{equation}
T^{\prime}(t)+\lambda^{2}T(t)=0.  \label{eq17}
\end{equation}

From (\ref{eq12})-(\ref{eq14}) we have the following boundary
conditions for $X_{j}(x)$:
\begin{equation}
X_{1}(0)= X_{2}(0)= X_{3}(0), \label{eq18}
\end{equation}
\begin{equation}
 \left.\left(X_{1}^{\prime}(x)+ X_{2}^{\prime}(x)+
X_{3}^{\prime}(x)\right)\right|_{x=0}=0, \label{eq19}
\end{equation}
\begin{equation}
X_{j}(L_{j})=0. \label{eq20}
\end{equation}

The general solution of equation (\ref{eq16}) can be written as
$$ X_{j}(x)=A_{j}\cos\frac{\lambda}{\kappa_j}(x-L_{j})+ B_{j}\sin\frac{\lambda}{\kappa_j}(x-L_{j})$$
From (\ref{eq20}) we have $ A_{j}=0, $ where the boundary
conditions \re{eq18} -\re{eq20} lead to the following linear
algebraic system (with respect to $B_{j}, j=1,2,3$):
$$ -B_{1}\sin\frac{\lambda}{\kappa_1}L_{1}+ B_{2}\sin\frac{\lambda}{\kappa_2}L_{2}=0, $$
$$ -B_{1}\sin\frac{\lambda}{\kappa_1}L_{1}+ B_{3}\sin\frac{\lambda}{\kappa_3}L_{3}=0, $$
$$ \frac{\lambda}{\kappa_1}B_{1}\cos\frac{\lambda}{\kappa_1}L_{1}+ \frac{\lambda}{\kappa_2}B_{2} \cos\frac{\lambda}{\kappa_2}L_{2}+ \frac{\lambda}{\kappa_3}B_{3} \cos\frac{\lambda}{\kappa_3}L_{3} =0, $$
which gives the following secular equation for finding the
eigenvalues, $\lambda_n$(provided $L_{1}, L_{2}, L_{3}$ are
rationally independent):
\begin{equation}
 \sum_{j=1}^{3}\frac{1}{\kappa_j}\cot\frac{\lambda}{\kappa_j}L_{j}=0.
 \label{eq21}
\end{equation}

\begin{figure}[th!]
\includegraphics[width=70mm]{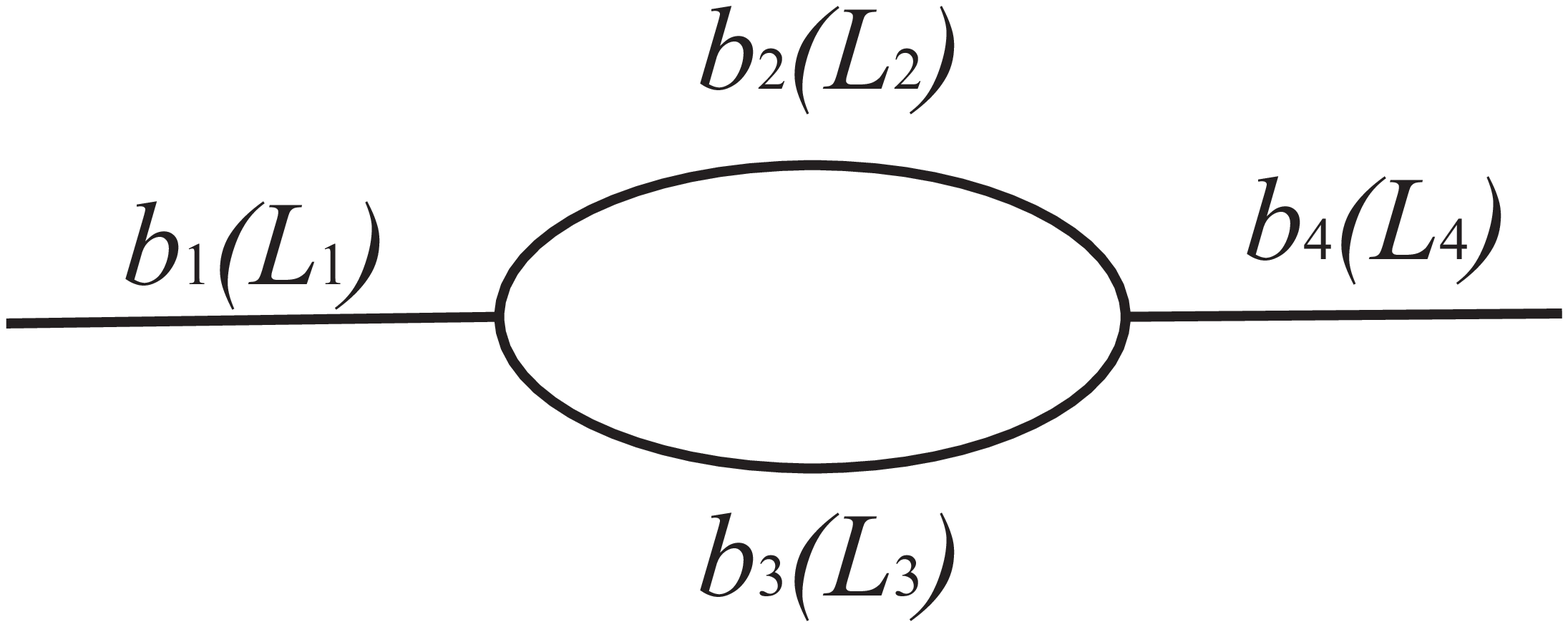}
\caption{Loop graph. $x_{i} \in [0,L_i]$ with $i=1,2,3,4$.}
\label{loop}
\end{figure}

Existence of the roots of Eq.\re{eq21} (which can be easily shown,
e.g., using the Newton's method) implies solvability of
Eq.\re{eq21} on metric graph. Therefore  finding numerically the
roots of Eq.\re{eq21} one can construct complete set of
eigenvalues of the problem given by Eqs.\re{eq10}-\re{eq14}.

Then the general solution of the heat equation on metric star
graph, which describes evolution of the temperature profile, can
be written as
\begin{equation}
 u_{j}(x,t)=\sum_{n=1}^{+\infty}a_{n}^{(j)}
e^{-\lambda_{n}^{2}t}\sin\frac{\lambda_n}{\kappa_j}(x-L_{j}).
 \label{eq24}
\end{equation}
where $ a_{n}^{(j)}= \underset{0}{\overset{L_j}{\int}}f_{j}(x)
\sin\frac{\lambda_n}{\kappa_j}(x-L_{j}) dx $.

Fig. \ref{tprof1}  provides evolution of the temperature profile
in metric star graph calculated using the solution of Eq.\re{eq24}
for the values of heat conductivity $\kappa_1=0.3,\,
\kappa_2=0.5,\, \kappa_3=0.7$ and bond lengths  $L_1=10,\,
L_2=11,\, L_3=12.$ The initial condition, i.e. the temperature
profile at $t=0$ is chosen as $f_j(x)=\text{sech}(x)$. An
important characteristics of the heat diffusion is the heat flux
determined as $Q_j(x,t) = -\kappa_j \frac{\partial u_j}{\partial
x}$.  Fig. \ref{hf1} presents plots of the heat flux vs coordinate
and time on each bond of a star graph for the values of heat
conductivity $\kappa_1=4.5,\, \kappa_2=8,\, \kappa_3=8.4$ and for
the same initial condition and  bond lengths as those in Fig.
\ref{tprof1}.

\section{Extension to other graphs}
The above treatment can be easily extended to other graph
topologies, such as tree and loop graphs. The tree graph (see Fig.
\ref{tree}) consists of three subgraphs $b_1, (b_{1i}),
(b_{1ij})$, where $i,j$ run over the given bonds of a subgraph. On
each bond $b_1, b_{1i}, b_{1ij}$ we have  heat equation given by
\be \frac{\partial u_{b}}{\partial t}=\kappa_b^2\frac{\partial^{2}
u_{b}}{\partial x^{2}}, \,\kappa_b>0 \label{he2}\ee for which the
initial and boundary conditions can be written as \be
u_b(x,0)=f_b(x),\,0<x<L_b, \ee \be u_{1}(0,t)=0,
u_{1ij}(L_{1ij},t)=0, i=1,2; j=1,2 \ee

\begin{figure}[th!]
\includegraphics[width=90mm]{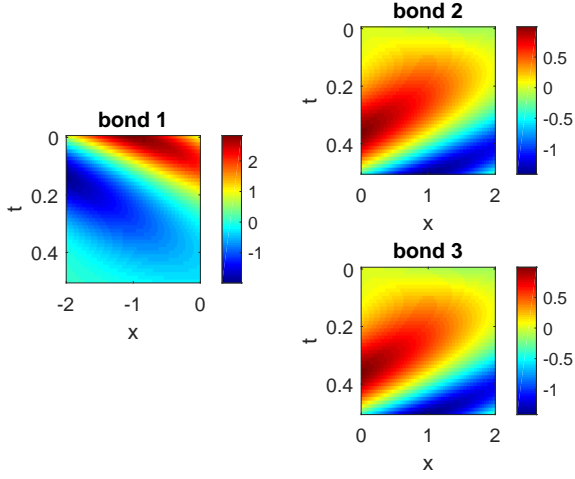}
\vspace{1mm} \caption{(Color online) Evolution of the temperature
profile plotted using the solution of the problem
(\ref{nlheq})-(\ref{bc1}) for  $\beta_1=1$, $\beta_2=\sqrt{2}$,
$\beta_3=\sqrt{2}$ and  $L_1=L_2=L_3=2$.} \label{tprof2}
\end{figure}

\be u_{1}(L_{1},t)=u_{1i}(0,t), i=1,2 \ee \be
u_{1i}(L_{1i},t)=u_{1ij}(0,t), i=1,2; j=1,2 \ee \be
\left.\frac{\partial u_{1}}{\partial x}\right|_{x=L_{1}}-
\sum_{i=1}^{2}\left.\frac{\partial u_{1i}}{\partial
x}\right|_{x=0}=0, i=1,2 \ee \be \left.\frac{\partial
u_{1i}}{\partial x}\right|_{x=L_{1i}}-
\sum_{j=1}^{2}\left.\frac{\partial u_{1ij}}{\partial
x}\right|_{x_{j}=0}=0, i=1,2. \ee

Variables in Eq.\re{he2} can be separated and we get from the
boundary conditions, the secular equation for finding the
eigenvalues, $\lambda_n$ that allows to obtain the complete set of
the eigenvalues and eigenfunctions similarly to that for the star
graph.

For the loop graph presented in Fig. \ref{loop}, we have
\begin{equation}
\frac{\partial}{\partial
t}u_{j}=\kappa_j^2\frac{\partial^{2}}{\partial
x^{2}}u_{j},\,\kappa_j>0 \label{eq25}
\end{equation}
\begin{equation}
u_{j}(x,0)=f_{j}(x) \label{eq26}
\end{equation}
\begin{equation}
u_{1}(0,t)=u_{4}(L_{4},t)=0 \label{eq27}
\end{equation}
\begin{equation}
u_{1}(L_{1},t)=u_{2}(0,t)=u_{3}(0,t) \label{eq28}
\end{equation}
\begin{equation}
u_{2}(L_{2},t)=u_{3}(L_{3},t)=u_{4}(0,t) \label{eq29}
\end{equation}
\begin{equation}
\left.\frac{\partial u_{1}}{\partial x}\right|_{x=L_{1}}
-\left.\frac{\partial u_{2}}{\partial x}\right|_{x=0}
-\left.\frac{\partial u_{3}}{\partial x}\right|_{x=0}=0.
\label{eq30}
\end{equation}
\begin{equation}
\left.\frac{\partial u_{2}}{\partial x}\right|_{x=L_{2}}
+\left.\frac{\partial u_{3}}{\partial x}\right|_{x=L_{3}}
-\left.\frac{\partial u_{4}}{\partial x}\right|_{x=0}=0.
\label{eq31}
\end{equation}


Separating variables by substitution \re{eq15} we get
Eqs.\re{eq16} and \re{eq17} with the boundary conditions given by
\begin{equation}
X_{1}(0)=X_{4}(L_{4})=0.
 \label{eq35}
\end{equation}
\begin{equation}
X_{1}(L_{1})=X_{2}(0)=X_{3}(0).
 \label{eq36}
\end{equation}
\begin{equation}
X_{2}(L_{2})=X_{3}(L_{3})=X_{4}(0).
 \label{eq37}
\end{equation}
\begin{equation}
\left.X_{1}'(x)\right|_{x=L_{1}}-\left.X_{2}'(x)\right|_{x=0}-\left.X_{3}'(x)\right|_{x=0}=0.
 \label{eq38}
\end{equation}
\begin{equation}
\left.X_{2}'(x)\right|_{x=L_{2}}+\left.X_{3}'(x)\right|_{x=L_{3}}-\left.X_{4}'(x)\right|_{x=0}=0.
 \label{eq39}
\end{equation}
General solution can be written as
$$
X_{j}(x)=A_{j}\cos\frac{\lambda}{\kappa_j}(x-L_{j})+B_{j}\sin\frac{\lambda}{\kappa_j}(x-L_{j}),
j={1,2,3,4} $$

After some algebra we get the following system of equations from
which we can find $\lambda_n$, $A_j$ and $B_j$: \vskip 1cm

$$
B_{1}\sin\frac{\lambda}{\kappa_1}L_{1}-A_{2}\cos\frac{\lambda}{\kappa_2}L_{2}-B_{2}\sin\frac{\lambda}{\kappa_2}L_{2}=0
$$
$$
B_{1}\sin\frac{\lambda}{\kappa_1}L_{1}-A_{3}\cos\frac{\lambda}{\kappa_3}L_{3}-B_{3}\sin\frac{\lambda}{\kappa_3}L_{3}=0
$$
$$ A_{2}-A_{3}=0 $$
$$ A_{2}+B_{4}\sin\frac{\lambda}{\kappa_4}L_{4}=0 $$
\begin{eqnarray}
\frac{\lambda}{\kappa_1}B_{1}\cos\frac{\lambda}{\kappa_1}L_{1}-\frac{\lambda}{\kappa_2}A_{2}\sin\frac{\lambda}{\kappa_2}L_{2}-\frac{\lambda}{\kappa_2}B_{2}\cos\frac{\lambda}{\kappa_2}L_{2}\nonumber\\
-\frac{\lambda}{\kappa_3}A_{3}\sin\frac{\lambda}{\kappa_3}L_{3}-\frac{\lambda}{\kappa_3}B_{3}\cos\frac{\lambda}{\kappa_3}L_{3}=0\nonumber
\end{eqnarray}
$$ \frac{\lambda}{\kappa_2}B_{2}+\frac{\lambda}{\kappa_3}B_{3}-\frac{\lambda}{\kappa_4}B_{4}\cos\frac{\lambda}{\kappa_4}L_{4}=0$$

Eigenvalues, eigenfunctions and general solutions can be found
similarly to those for star and tree graphs.

\begin{figure}[th!]
\includegraphics[width=90mm]{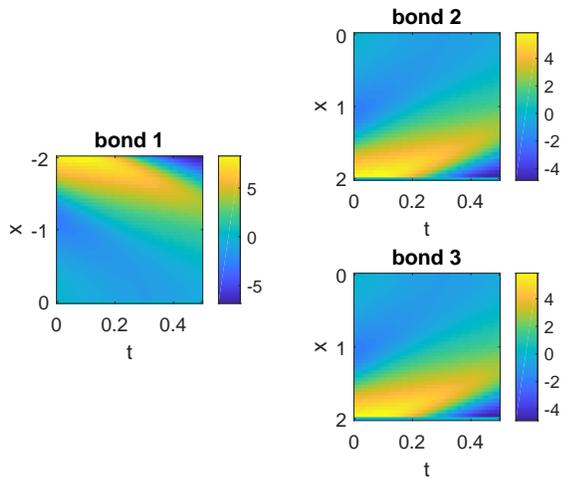}
\vspace{1mm} \caption{(Color online) Heat flux plotted using the
solution of the problem \re{nlheq}-\re{bc1} for $\beta_1=1$,
$\beta_2=\sqrt{2}$, $\beta_3=\sqrt{2}$ and $L_1=L_2=L_3=2$.}
\label{hf2}
\end{figure}

\section{Heat diffusion in nonlinear regime}

The above model for the linear heat diffusion in networks can be
extended to the case of nonlinear regime, using a nonlinear heat
equation on metric graph.  Here we demonstrate it for a star graph
with three semi-infinite bonds.  Consider the following nonlinear
heat equation on a star graph with the  bonds
$b_1\sim(-\infty;0),\,b_{2,3}\sim(0;+\infty)$:
\begin{equation}
\frac{\partial u_j}{\partial t}-\frac{\partial^2u_j}{\partial
x^2}=-2\beta_j^2u_j^3,\,x\in b_j,\,t>0\label{nlheq}
\end{equation}
for which the  vertex boundary conditions are imposed as
\begin{eqnarray}
&\beta_1u_1(0,t)=\beta_2u_2(0,t)=\beta_3u_3(0,t),\nonumber\\
&\left.\frac{1}{\beta_1}\frac{\partial u_1}{\partial
x}\right|_{x=0}=\left.\frac{1}{\beta_2}\frac{\partial
u_2}{\partial
x}\right|_{x=0}+\left.\frac{1}{\beta_3}\frac{\partial
u_3}{\partial x}\right|_{x=0}.\label{bc1}
\end{eqnarray}
Solution of the equation (\ref{nlheq}) (without boundary
conditions) can be written as \ci{NLHE1}
\begin{equation}
u_j(x,t)=\frac{x-x_0}{\beta_j}\text{sd}\left((x-x_0)^2+6t,\frac{1}{\sqrt{2}}\right).\label{sol1}
\end{equation}
Fulfilling the boundary conditions (\ref{bc1}) by this solution is
possible if coefficients  $\beta_j,\,j=1,2,3$ obey the following
(constraints) sum rule:
\begin{equation}
\frac{1}{\beta_1^2}=\frac{1}{\beta_2^2}+\frac{1}{\beta_3^2}.\label{constrain1}
\end{equation}

Eq.\re{constrain1} presents the condition for integrability of
nonlinear heat equation on metric star graph.  Fig.\ref{tprof2}
present evolution of the temperature profile plotted using
Eq.\re{sol1} for the case when the sum rule given by
Eq.\re{constrain1} is fulfilled. Corresponding plot of the heat
flux on each bond is presented in Fig.\ref{hf2}. Comparing the
plots of the heat flux with those for linear case we can conclude
that in nonlinear regime heat flux is more intensive than that in
linear one.
\section{Conclusions}
We studied heat diffusion on branched systems by considering
linear and nonlinear regimes of diffusion. Exact analytical
solutions of linear heat equation on simple metric graphs are
obtained. Temperature profiles for such systems are computed.
Nonlinear regime is studied on the basis of a nonlinear heat
equation on metric star graph. Exact solutions are obtained for
the case, when nonlinearity coefficient fulfills some constraint
which can be written in the form of sum rule. The above models can
be  applied for the study of linear and nonlinear heat diffusion
in different branched materials such as polymers, carbon nanotube
networks, DNA and different low-dimensional networks.

\end{document}